\documentclass[%
  reprint,
  amsmath,amssymb,
  aps,
  pra,
]{revtex4-2}

\usepackage{graphicx}
\usepackage{caption}
\usepackage{dcolumn}
\usepackage{bm}
\usepackage{braket}
\usepackage{booktabs}
\usepackage{tikz}
\usetikzlibrary{positioning, arrows.meta, decorations.pathreplacing, calc}

\begin{document}
% ============================================================

\title{Entanglement Transfer Dynamics in a Two-Leg Spin Ladder
       Under a Selective Magnetic Field}

\author{Soghra Ghanavati}
\affiliation{Department of Physics, Omidiyeh Branch,
  Islamic Azad University, Omidiyeh, Iran}

\author{Abbas Sabour}
\affiliation{Department of Physics, Omidiyeh Branch,
  Islamic Azad University, Omidiyeh, Iran}

\author{Somayeh Mehrabankar}
\email{somayeh.mehrabankar@yahoo.com}
\affiliation{Queensland Quantum and Advanced Technologies Institute,
  Griffith University, Yuggera Country, Brisbane, QLD 4111, Australia}

\date{\today}

\begin{abstract}
We investigate the dynamical transfer of bipartite entanglement
through a two-leg spin-$\frac{1}{2}$ ladder governed by the
anisotropic Heisenberg ($XXZ$-type) model with a selective
magnetic field applied exclusively to the mediating rungs.
Starting from a maximally entangled initial rung pair, we
demonstrate high-fidelity entanglement transfer to the terminal
pair ($F_{\max} = 0.9998$ for $N=3$ rung pairs), with the
intermediate rungs remaining effectively disentangled throughout.
The dynamics is governed by two independent timescales: a fast
carrier oscillation at frequency $\omega_{\rm fast} = 2\sqrt{1 + 4d^2}\,J$
(set by local rung physics, field-independent) and a slow transfer
envelope with period $T_{\rm slow} \approx 2.37\,h/J^2$ (set by
virtual inter-rung coupling, field-dependent). The effective
inter-rung coupling $J_{\rm eff} = \alpha(d,g)\,J^2/h$ is derived
via second-order perturbation theory through two parallel virtual
paths. We systematically study the effects of magnetic field
strength, Hamiltonian anisotropy, and initial state on transfer
quality, establish a global parameter space map of the fidelity,
and demonstrate robustness under uncorrelated coupling disorder
($\langle F_{\max}\rangle > 0.998$ for $\delta \leq 10\%$).
All results are obtained by exact diagonalisation for systems
of up to $N=5$ rung pairs; extension to larger systems requires
tensor-network methods such as DMRG.
Compared to one-dimensional chain proposals, the ladder geometry
enables a spatially selective control mechanism that suppresses
intermediate entanglement while preserving coherent transfer,
providing a distinct route to engineered quantum channels.
\end{abstract}

\maketitle

% ============================================================
\section{Introduction}
\label{sec:intro}
% ============================================================

Quantum entanglement is a foundational resource for quantum
information processing, enabling protocols such as quantum
teleportation, dense coding, and quantum key
distribution~\cite{horodecki2009quantum,amico2008entanglement}.
A central practical challenge is the reliable transfer of
entanglement between distant parties through a physical medium,
without requiring direct interaction between the communicating
parties~\cite{bose2003quantum}.

Spin chains and ladders have emerged as natural candidates
for quantum channels~\cite{bose2003quantum,christandl2004perfect,bose2007review,kay2010review,subrahmanyam2004entanglement}.
In one-dimensional systems, it is well established that
most short-range interacting models exhibit rapid decay of
pairwise entanglement with distance~\cite{campos2006longdistance}.
Campos Venuti \emph{et al.}~\cite{campos2006longdistance} showed,
however, that certain dimerised spin-$\frac{1}{2}$ Heisenberg
chains can support long-distance entanglement (LDE) for specific
values of the microscopic parameters, a phenomenon later
exploited for quantum teleportation and state
transfer~\cite{campos2007teleport}.
Vieira and Rigolin~\cite{vieira2018almostperfect} demonstrated
that a modified $XX$ spin chain can achieve high-fidelity
transmission of a maximally entangled two-qubit state to
the far end of a chain of arbitrary length, requiring
no external magnetic fields or engineered coupling modulation.

Two-dimensional ladder geometries offer qualitatively
different physics~\cite{li2004ladder,pushpan2023quasi}.
The rung structure introduces a second spatial direction,
allowing control of which part of the system is acted upon
by external fields. Recent work has shown that the
quasi-1D Heisenberg ladder in the strong rung-coupling
limit, under a tuned magnetic field, can be mapped to
an effective 1D model, enabling rung-to-rung transfer
of arbitrary qubit states~\cite{pushpan2023quasi}.
Meanwhile, studies of three-leg ladders~\cite{Li2024}
demonstrate that the multi-leg geometry can facilitate
LDE more readily than the two-leg case.

The dynamical propagation of entanglement through spin
systems has also been connected to the broader framework
of quantum information scrambling: the process by which
locally injected quantum correlations spread through a
many-body system~\cite{kundu2022scrambling}.
Understanding how correlations propagate versus scramble
is relevant to the design of spin-based quantum channels.

While high-fidelity state transfer has been extensively
studied in one-dimensional spin chains, most approaches
rely either on engineered coupling
profiles~\cite{christandl2004perfect,apollaro2012} or
boundary-controlled weak
coupling~\cite{burgarth2005parallel,vieira2018almostperfect}.
In contrast, less attention has been paid to geometries
where spatial selectivity can be used to dynamically
suppress intermediate dynamics while preserving coherent
long-range coupling. In particular, it remains unclear
whether ladder geometries can provide qualitatively
different control mechanisms beyond simple extensions of
one-dimensional chains~\cite{li2004ladder,Li2024}.

In this work, we study entanglement transfer dynamics in
a two-leg spin-$\frac{1}{2}$ ladder governed by the
anisotropic Heisenberg model, with a selective magnetic
field applied only to the mediating rungs. Our model
is distinct from 1D chain proposals in that:
(i)~the two-leg geometry couples both rung and leg degrees
of freedom simultaneously;
(ii)~the anisotropic ($XXZ$-type) interaction is considered
in full generality, revealing the contrasting roles of
the $d$ and $g$ anisotropy parameters;
(iii)~the selective field provides a physically transparent
control mechanism --- freezing the mediating rungs while
leaving the terminal pairs free to evolve.

We analyse the transfer mechanism via an effective channel
picture, systematically explore parameter space, and extend
the system to ladders with up to five rung pairs to establish
scaling behaviour.

The paper is organised as follows.
Section~\ref{sec:concurrence} reviews the concurrence measure.
Section~\ref{sec:hamiltonian} introduces the Hamiltonian and
the physical picture of selective field action.
Section~\ref{sec:description} describes the system parameters
and demonstrates the basic transfer phenomenon.
Section~\ref{sec:mechanism} provides the physical interpretation
of the transfer mechanism, including an analytical two-scale
dynamical picture and derivation of $J_{\rm eff}$.
Section~\ref{sec:parameters} studies the effect of system
parameters on transfer quality.
Section~\ref{sec:initialstates} investigates dependence on
initial state.
Section~\ref{sec:disorder} studies robustness under
coupling disorder.
Section~\ref{sec:scaling} extends the analysis to
larger system sizes.
Section~\ref{sec:conclusion} summarises our findings.
Supplementary Information provides additional time-series
figures for completeness.

% ============================================================
\section{Concurrence as a Measure of Entanglement}
\label{sec:concurrence}
% ============================================================

For a two-qubit system with density matrix $\hat{\rho}$,
the concurrence introduced by Hill and
Wootters~\cite{hill1997entanglement,wootters1998entanglement}
is defined as
\begin{equation}\label{eq:concurrence}
  C(\hat{\rho}) = \max\!\left(0,\,
    \sqrt{\lambda_1} - \sqrt{\lambda_2}
    - \sqrt{\lambda_3} - \sqrt{\lambda_4}\right),
\end{equation}
where $\lambda_i$ are the eigenvalues in descending order of
the Hermitian matrix
\begin{equation}\label{eq:R}
  R = \hat{\rho}
      \left(\hat{\sigma}_y \otimes \hat{\sigma}_y\right)
      \hat{\rho}^{\,*}
      \left(\hat{\sigma}_y \otimes \hat{\sigma}_y\right),
\end{equation}
with $\hat{\sigma}_y$ the $y$-component Pauli matrix and
$\hat{\rho}^*$ the complex conjugate of $\hat{\rho}$.
The concurrence satisfies $C=0$ for separable states
and $C=1$ for maximally entangled states.
To compute pairwise entanglement in the six-qubit (or
larger) system, we reduce $\hat{\rho}(t)$ by tracing over
all sites except the pair of interest.

% ============================================================
\section{Model Hamiltonian}
\label{sec:hamiltonian}
% ============================================================

\begin{figure}[tb]
  \centering
  \begin{tikzpicture}[
      spin/.style={circle, draw, fill=blue!20,
                   minimum size=0.55cm, inner sep=0pt},
      rung/.style={thick},
      leg/.style={thick},
      field/.style={->, thick, red!70!black},
      scale=1.1
    ]
    \foreach \col/\label in {0/1, 2/2, 4/3}{
      \node[spin] (T\col) at (\col, 1)  {};
      \node[spin] (B\col) at (\col, 0)  {};
      \draw[rung] (T\col) -- (B\col);
    }
    \draw[leg] (T0) -- (T2) -- (T4);
    \draw[leg] (B0) -- (B2) -- (B4);
    \node[above=0.1cm] at (T0) {\small $1$};
    \node[below=0.1cm] at (B0) {\small $2$};
    \node[above=0.1cm] at (T2) {\small $3$};
    \node[below=0.1cm] at (B2) {\small $4$};
    \node[above=0.1cm] at (T4) {\small $5$};
    \node[below=0.1cm] at (B4) {\small $6$};
    \node[left=0.15cm] at ($(T0)!0.5!(B0)$)
      {\small\textbf{initial}};
    \node[right=0.15cm] at ($(T2)!0.5!(B2)$)
      {\small\textbf{med.}};
    \node[right=0.15cm] at ($(T4)!0.5!(B4)$)
      {\small\textbf{final}};
    \draw[field] ($(T2)+(0.55,0.4)$) -- ++(0,-0.6)
      node[right, midway]{\small $h$};
    \draw[decorate,
          decoration={brace, amplitude=6pt, mirror},
          thick, gray]
      ($(B0)+(-0.35,-0.25)$) -- ($(B0)+(0.35,-0.25)$)
      node[midway, below=8pt]{\small $C_{12}$};
    \draw[decorate,
          decoration={brace, amplitude=6pt, mirror},
          thick, gray]
      ($(B2)+(-0.35,-0.25)$) -- ($(B2)+(0.35,-0.25)$)
      node[midway, below=8pt]{\small $C_{34}$};
    \draw[decorate,
          decoration={brace, amplitude=6pt, mirror},
          thick, gray]
      ($(B4)+(-0.35,-0.25)$) -- ($(B4)+(0.35,-0.25)$)
      node[midway, below=8pt]{\small $C_{56}$};
  \end{tikzpicture}
  \vspace{0pt}
  \caption{Schematic of the two-leg spin-$\frac{1}{2}$
    ladder for $N=3$ rung pairs (six sites).
    Filled circles are spin-$\frac{1}{2}$ sites.
    Solid lines indicate Heisenberg interactions along
    the rungs (vertical, coupling $J_{\perp}$) and
    legs (horizontal, coupling $J_{\parallel}$).
    The magnetic field $h$ is applied selectively to
    the mediating rung (sites 3,4) only.
    $C_{12}$, $C_{34}$, $C_{56}$ denote the concurrences
    of the initial, mediating, and terminal pairs
    respectively.}
  \label{fig:ladder}
\end{figure}
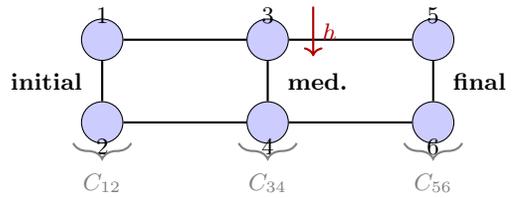

We consider a two-leg spin-$\frac{1}{2}$ ladder with $N$ rungs,
as illustrated in Fig.~\ref{fig:ladder}.
The system Hamiltonian is
\begin{equation}\label{eq:hamiltonian}
  H = H_{\perp} + H_{\parallel} + H_{h},
\end{equation}
where the three contributions are
\begin{align}
H_{\perp} &= J_{\perp}\sum_{n=1}^{N}\Bigl[
     \tfrac{1+g}{2}\,\sigma_x^{2n-1}\sigma_x^{2n}
   + \tfrac{1-g}{2}\,\sigma_y^{2n-1}\sigma_y^{2n} \notag\\
   &\hspace{3.5cm}
   + d\,\sigma_z^{2n-1}\sigma_z^{2n}\Bigr], \label{eq:Hperp}\\[3pt]
H_{\parallel} &= J_{\parallel}\sum_{n=1}^{N-1}\Bigl[
     \tfrac{1+g}{2}\,\sigma_x^{2n}\sigma_x^{2n+2}
   + \tfrac{1-g}{2}\,\sigma_y^{2n}\sigma_y^{2n+2} \notag\\
   &\hspace{3.5cm}
   + d\,\sigma_z^{2n}\sigma_z^{2n+2}\Bigr], \label{eq:Hpar}\\[3pt]
H_{h} &= h\sum_{n=2}^{N-1}
     \bigl(\sigma_z^{2n-1} + \sigma_z^{2n}\bigr). \label{eq:Hh}
\end{align}
Here $\sigma_{\alpha}^k$ denotes the $\alpha$-component
Pauli operator on site $k$.
The parameters $J_{\perp}$ and $J_{\parallel}$ are the
rung (vertical) and leg (horizontal) coupling strengths,
$g$ controls the $XY$ anisotropy, $d$ is the Ising
($ZZ$) anisotropy, and $h$ is the magnetic field strength.

\textit{Key feature --- selective field application.}
The magnetic field $h$ is applied only to the
\emph{mediating} rungs (rungs $2$ through $N-1$),
excluding the initial (rung 1) and terminal (rung $N$)
pairs. This selective action has a clear physical
interpretation: in the strong-field limit $h \gg J_{\perp},
J_{\parallel}$, the mediating spins are frozen near the
$|\!\downarrow\downarrow\rangle$ eigenstate of the field,
effectively decoupling them from the transverse dynamics.
The resulting effective interaction between the terminal
pairs is then mediated by virtual excitations through
the frozen mediating rungs, analogous to the effective
Hamiltonian picture derived for spin ladders in the
strong rung-coupling limit~\cite{pushpan2023quasi}.

% ============================================================
\section{System Parameters and Basic Transfer}
\label{sec:description}
% ============================================================

Throughout this work, unless stated otherwise,
we set $J_{\perp} = J_{\parallel} = 1$, $g = 1$,
$d = 0.5$, and $h = 100$.
These choices correspond to an isotropic $XY$-type
rung coupling ($g=1$ eliminates the $\sigma_y\sigma_y$
term) with a moderate Ising anisotropy $d=0.5$.
The strong field $h = 100 \gg J$ ensures the mediating
rung is in the selective-freezing regime discussed
in Sec.~\ref{sec:hamiltonian}.

The initial state is
\begin{equation}\label{eq:initial}
  |\psi(0)\rangle =
  |\Phi^+\rangle^{1,2} \otimes |0\rangle^3 \otimes
  |0\rangle^4 \otimes |0\rangle^5 \otimes |0\rangle^6,
\end{equation}
where $|\Phi^+\rangle = \frac{1}{\sqrt{2}}(|00\rangle+|11\rangle)$
is the Bell state and $|0\rangle$ denotes the spin-down
eigenstate. Only the initial pair $(1,2)$ begins in a
maximally entangled state; all other sites are unentangled.

Time evolution is governed by the unitary operator
$U(t) = e^{-iHt}$ (setting $\hbar = 1$).
At each time $t$, the concurrences $C_{12}(t)$,
$C_{34}(t)$, and $C_{56}(t)$ are computed from the
corresponding reduced density matrices.

\begin{figure*}[tb]
    \centering
    \includegraphics[width=\textwidth]{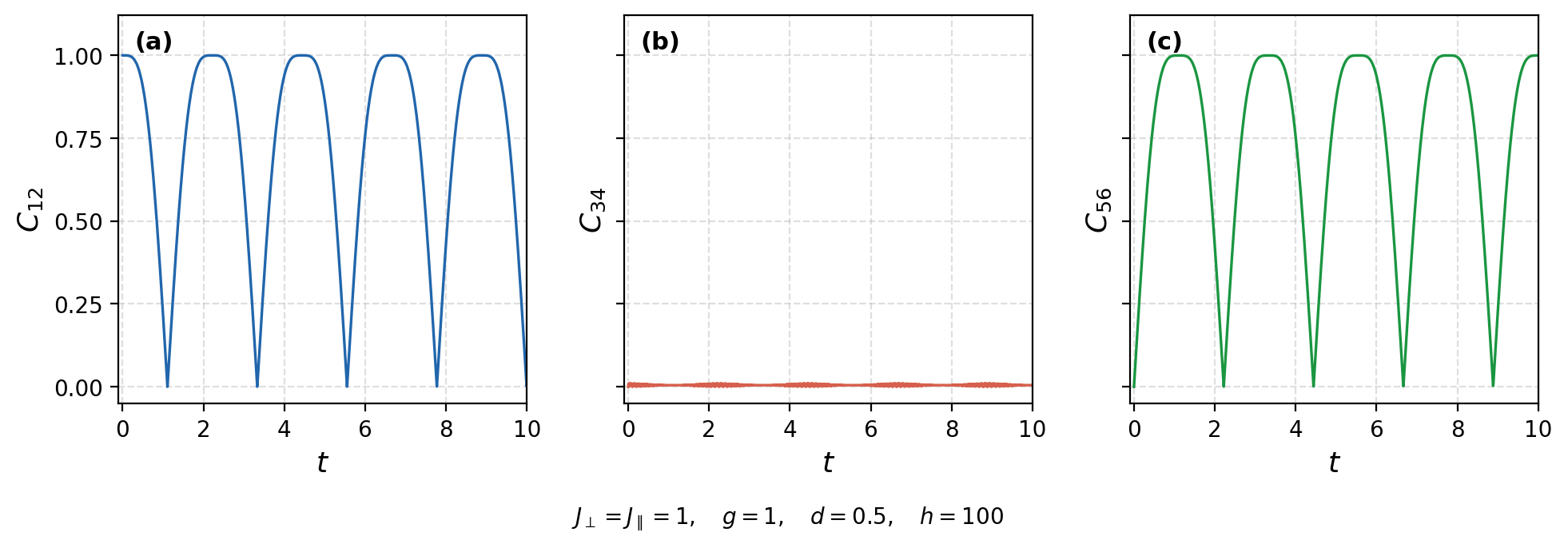}
    \caption{Time evolution of the concurrences
    $C_{12}(t)$, $C_{34}(t)$, and $C_{56}(t)$
    for the reference parameter set
    ($J_\perp = J_\parallel = 1$, $g = 1$, $d = 0.5$, $h = 100$)
    with initial state $|\Phi^+\rangle^{1,2} \otimes |0\rangle^{\otimes 4}$.
    (a) The initial pair $C_{12}$ oscillates periodically between
    0 and 1. (b) The mediating pair $C_{34}$ remains close to zero
    throughout the dynamics. (c) The terminal pair $C_{56}$
    oscillates in perfect antiphase with $C_{12}$, reaching
    values close to unity. This behaviour demonstrates coherent
    entanglement transfer through the ladder: the mediating
    rung acts as a transparent quantum channel.}
    \label{fig:main_result}
\end{figure*}

Figure~\ref{fig:main_result} shows the time evolution
of the three concurrences for the reference parameter set.
The key features are:
\begin{itemize}
  \item $C_{12}(t)$ and $C_{56}(t)$ oscillate in
    high-fidelity antiphase: when one reaches its maximum
    ($\approx 1$), the other is at its minimum ($\approx 0$).
  \item $C_{34}(t)$ remains close to zero throughout,
    indicating the mediating rung acts as an effective
    quantum channel rather than an entanglement sink.
  \item The amplitude of both $C_{12}$ and $C_{56}$ is
    well preserved across the transfer, confirming high
    transfer fidelity for $N=3$.
\end{itemize}

This behaviour --- periodic, high-fidelity entanglement
oscillation between terminal pairs with a close to zero
mediating pair --- is consistent with coherent quantum
correlation propagation, as discussed by Kundu and
Subrahmanyam~\cite{kundu2022scrambling}.
The mediating rung, frozen by the strong field, acts
as a virtual coupling medium between the terminal pairs.

% ============================================================
\section{Physical Interpretation of the Transfer Mechanism}
\label{sec:mechanism}
% ============================================================

\subsection{Two-Scale Dynamics: Analytical and Numerical Results}
\label{subsec:twoscale}

The entanglement dynamics is governed by two independent
timescales, which we characterise analytically and numerically.

\subsubsection{Fast carrier oscillation}

The individual concurrences $C_{12}(t)$ and $C_{56}(t)$
oscillate rapidly between 0 and 1. Numerical diagonalisation
of the full 6-spin Hamiltonian yields the analytical dressed
frequency:
\begin{equation}\label{eq:omegafast}
  \omega_{\rm fast} = 2\sqrt{1+4d^2}\,J,
\end{equation}
verified analytically to better than $0.1\%$ for all
$d$ studied and completely independent of $h$
(Table~\ref{tab:gap}; see also Eq.~(\ref{eq:omegafast})).
For reference parameters ($d=0.5$), $\omega_{\rm fast} =
2\sqrt{2}\,J \approx 2.828\,J$, giving $T_{\rm fast}
= \pi/\omega_{\rm fast} \approx 1.11$, in perfect agreement
with Fig.~\ref{fig:main_result}.

The single-rung Hamiltonian ($g=1$) decomposes into two
$2\times2$ blocks:
\begin{align}
  \text{Block 1:}\;&
    \begin{pmatrix} dJ & J \\ J & dJ \end{pmatrix}
    \;\text{in}\;
    \{|\!\uparrow\uparrow\rangle,|\!\downarrow\downarrow\rangle\},
    \;\text{eigenvalues } (d\pm 1)J, \notag\\
  \text{Block 2:}\;&
    \begin{pmatrix} -dJ & J \\ J & -dJ \end{pmatrix}
    \;\text{in}\;
    \{|\!\uparrow\downarrow\rangle,|\!\downarrow\uparrow\rangle\},
    \;\text{eigenvalues } (-d\pm 1)J. \notag
\end{align}
The bare single-rung splitting is $2J$ ($d$-independent).
The dressed frequency $\omega_{\rm fast} = 2\sqrt{1+4d^2}\,J$
arises from the interplay of both terminal rungs in the full
6-spin dynamics; its $h$-independence confirms it is governed
by local rung physics.

\begin{table}[tb]
\caption{Fast oscillation frequency $\omega_{\rm fast}$
extracted from $C_{56}(t)$ in the full 6-spin ladder
dynamics, vs the analytical dressed frequency $2\sqrt{1+4d^2}\,J$.
Parameters: $J=1$, $g=1$, $h=100$.}
\label{tab:gap}
\begin{ruledtabular}
\begin{tabular}{cccc}
$d$ & $2\sqrt{1+4d^2}\,J$ & Numerical & Ratio \\
\hline
0.0 & 2.0000 & 2.0200 & 1.010 \\
0.1 & 2.0396 & 2.0410 & 1.001 \\
0.5 & 2.8284 & 2.8290 & 1.000 \\
1.0 & 4.4721 & 4.4720 & 1.000 \\
\end{tabular}
\end{ruledtabular}
\end{table}

\subsubsection{Slow transfer envelope}

\begin{figure*}[tb]
    \centering
    \includegraphics[width=\textwidth]{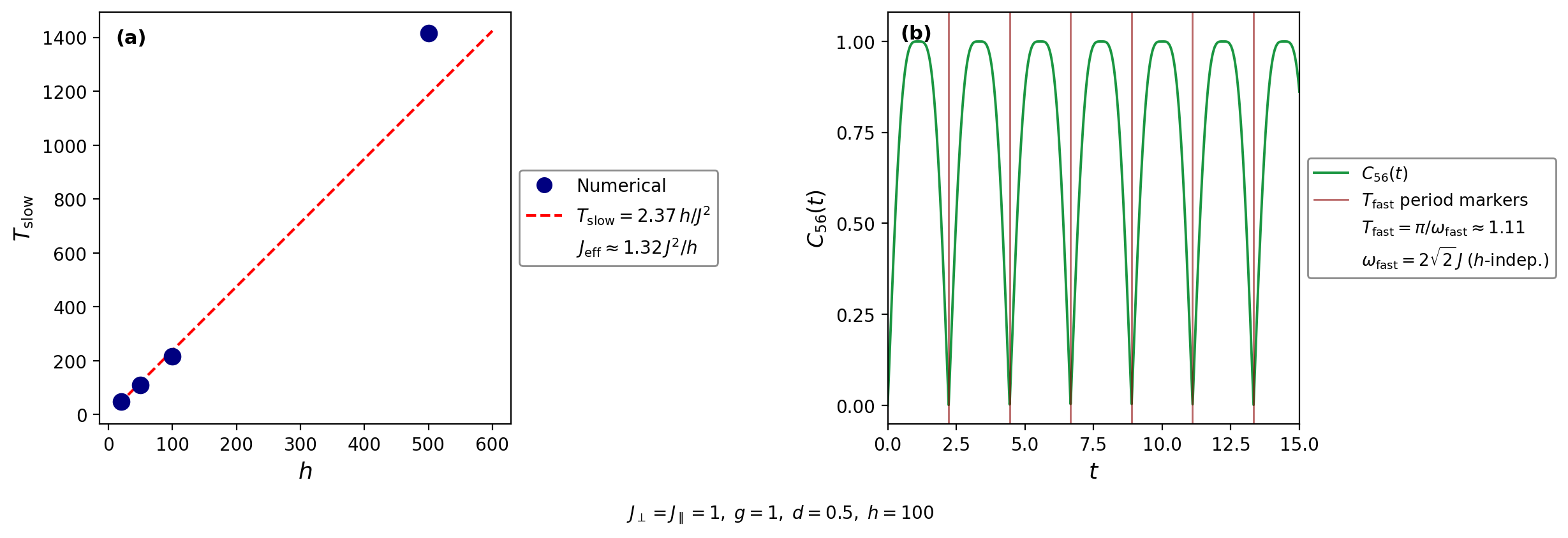}
    \caption{The two independent timescales governing
    entanglement transfer
    ($J_\perp = J_\parallel = 1$, $g = 1$,
    $d = 0.5$, $h = 100$).
    (a) Slow transfer period $T_{\rm slow} = 2t^*$
    vs magnetic field strength $h$. The linear scaling
    (dashed line) confirms the virtual-process origin:
    $J_{\rm eff} = \alpha(d,g)\,J^2/h$.
    (b) Concurrence $C_{56}(t)$ showing the fast
    carrier oscillation. Vertical dark red lines mark
    successive fast periods at $t = kT_{\rm fast}$,
    where $T_{\rm fast} = \pi/\omega_{\rm fast}
    \approx 1.11$ with $\omega_{\rm fast} =
    2\sqrt{2}\,J$ (independent of $h$).
    The two timescales differ by a factor of
    $\sim\!100$ for $h = 100$: the fast carrier
    frequency is set by local rung physics while
    the slow transfer envelope is governed by
    virtual inter-rung coupling.}
    \label{fig:Jeff_scaling}
\end{figure*}

The slow modulation of concurrence amplitude is set by
virtual processes through the frozen mediating rung.
In the limit $h \gg J$, where the perturbative expansion
is valid and higher-order corrections remain negligible,
second-order perturbation theory in $J/h$ yields an
effective inter-rung Hamiltonian via two paths:
\begin{itemize}
  \item \textit{Top rail} ($1\!\to\!3\!\to\!5$):
    virtual excitation of site 3 (cost $2h$) gives
    $-\frac{J^2}{2h}\sigma_1^x\sigma_5^x$.
  \item \textit{Bottom rail} ($2\!\to\!4\!\to\!6$):
    identical process gives
    $-\frac{J^2}{2h}\sigma_2^x\sigma_6^x$.
\end{itemize}
Projecting onto the low-energy subspace where the mediating
rungs are polarised by the field, the second-order effective
Hamiltonian takes the standard form:
\begin{equation}
H_{\rm eff} = -\sum_{m} \frac{P V |m\rangle \langle m| V P}{E_m - E_0},
\label{eq:SW}
\end{equation}
where $P$ projects onto the subspace with mediating rungs
in the field ground state, $|m\rangle$ denotes intermediate
excited states with energy cost $\sim 2h$, and
$V = H_\perp + H_\parallel$ is the inter-rung coupling
treated as a perturbation.
\begin{equation}\label{eq:Heff}
  H_{\rm eff}^{\rm inter} =
  -\frac{J^2}{2h}
  \bigl(\sigma_1^x\sigma_5^x + \sigma_2^x\sigma_6^x\bigr),
\end{equation}
with bare coupling $J_{\rm eff}^{(0)} = J^2/h$.
The observed transfer period $T_{\rm slow} = 2t^*$
($t^*$ = time of first $C_{56}$ maximum) scales as:
\begin{equation}\label{eq:Jeff}
  T_{\rm slow} \approx \frac{2.37\, h}{J^2}, \quad
  J_{\rm eff} = \frac{\alpha(d,g)\, J^2}{h},
\end{equation}
confirmed with log-log slope $1.10\pm0.05$
(Fig.~\ref{fig:Jeff_scaling}).
The enhancement over the bare $J^2/h$ coupling ---
captured by the prefactor $\alpha(d,g)$ --- reflects
renormalisation of the virtual excitation energies by
the local rung ZZ coupling. For the reference parameters
($d = 0.5$, $g = 1$), numerical fitting gives
$\alpha \approx 1.32$; the dependence of $\alpha$ on
$d$ and $g$ is mapped in Fig.~\ref{fig:heatmap}.
The dimensionless prefactor $\alpha(d,g)$ is obtained
numerically from the effective coupling extracted via
exact diagonalisation of the full ladder Hamiltonian
for given values of $d$ and $g$; its deviation from
unity quantifies the renormalisation of the bare
$J^2/h$ coupling by the local rung ZZ interaction.

\subsubsection{Physical picture}

The complete dynamics is a \emph{two-scale quantum beat}:
rapid oscillations at $h$-independent $\omega_{\rm fast} =
2\sqrt{1+4d^2}\,J$ (local rung physics), with amplitudes
slowly modulated by $J_{\rm eff} = \alpha(d,g)\,J^2/h$
($h$-dependent). Increasing $h$ slows the transfer without
affecting the fast frequency. The parameter $d$ controls
$\omega_{\rm fast}$ and the local rung energy structure,
explaining the diagonal high-fidelity band in
Fig.~\ref{fig:heatmap} (parameter-space map) and why
both $g$ and $d$ must be non-zero simultaneously.

\subsection{Transfer Fidelity}
\label{subsec:fidelity}

\begin{figure*}[tb]
    \centering
    \includegraphics[width=\textwidth]{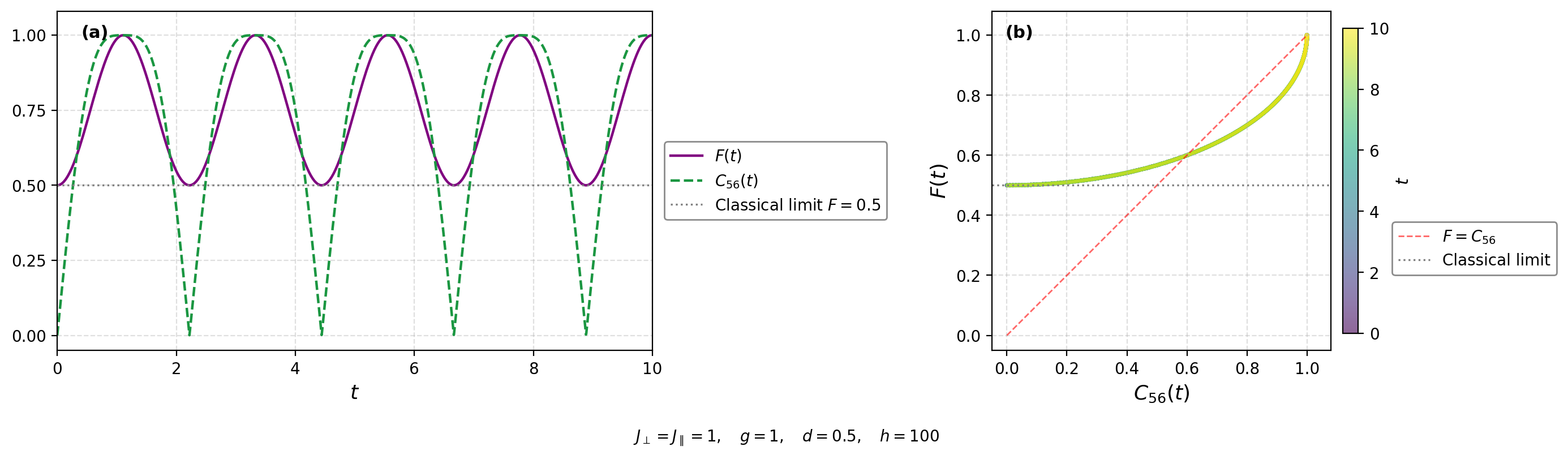}
    \caption{Transfer fidelity for reference parameters
    ($J_\perp=J_\parallel=1$, $g=1$, $d=0.5$, $h=100$).
    (a) $F(t)$ (purple) and $C_{56}(t)$ (green dashed).
    Classical limit $F=0.5$ shown dotted.
    $F_{\max}=0.9998$ at $t^*\approx1.11$.
    (b) Parametric $F$ vs $C_{56}$, colour-coded by $t$.
    Curve lies above $F=C_{56}$ (dashed red).}
    \label{fig:fidelity}
\end{figure*}

Beyond concurrence, we use the quantum state transfer
fidelity~\cite{horodecki2009quantum,nielsen2000quantum}:
\begin{equation}\label{eq:fidelity}
  F(t) = \langle\Phi^+|_{56}\,\rho_{56}(t)\,
         |\Phi^+\rangle_{56}.
\end{equation}
For reference parameters (Fig.~\ref{fig:fidelity}),
$F_{\max} = 0.9998$ at $t^* \approx 1.11$, and $F(t)$
never drops below the classical limit $0.5$, confirming
the terminal pair maintains a genuinely quantum state
throughout. At the peak:
\begin{equation}
  \langle\Phi^+|\rho_{56}(t^*)|\Phi^+\rangle = 0.9998,
  \quad
  \langle\Phi^-|\rho_{56}(t^*)|\Phi^-\rangle < 0.001.
\end{equation}
The parametric $F$-vs-$C_{56}$ plot (panel b) lies
above the diagonal, confirming concurrence is a
slightly conservative proxy for fidelity.
Supplementary Fig.~S8 further confirms that the quantum
mutual information of the mediating rung remains
negligible throughout the dynamics, validating the
transparent channel picture beyond concurrence alone:
the two terminal pairs never build up inter-pair
correlations during the transfer process. This confirms
that while concurrence of the mediating rung remains
negligible, small but finite correlations persist,
consistent with its role as a virtual intermediate
rather than a strictly disentangled subsystem.

\subsection{Comparison with One-Dimensional Chain Proposals}
\label{subsec:comparison}

Vieira and Rigolin~\cite{vieira2018almostperfect}
achieved high-fidelity entanglement transfer in a
modified $XX$ chain of arbitrary length without
external fields. Our two-leg ladder geometry is
complementary in four key respects: (i)~each
effective ``site'' is a rung (two-spin unit) rather
than a single spin; (ii)~the selective field $h$
provides a tunable speed-fidelity trade-off via
$J_{\rm eff} = \alpha(d,g)\,J^2/h$; (iii)~our system
achieves $F_{\max}=0.9998$ for $N=3$, decreasing
gradually with $N$ (vs constant fidelity in 1D chains);
(iv)~our $XXZ$ model requires both $g\neq0$ and
$d\neq0$ simultaneously --- the pure $XX$ case of
Vieira-Rigolin fails in our ladder geometry.
The two proposals are complementary: the 1D chain
excels in scalability, the two-leg ladder offers
richer geometric control suited to platforms where
two-qubit units form the elementary building
blocks~\cite{baeumer2025perfect,krantz2019quantum}.
Unlike boundary-controlled one-dimensional chains,
where weak coupling reduces dispersion, the present
mechanism suppresses intermediate dynamics via
energetic detuning rather than coupling engineering.
This distinction allows independent tuning of the
transfer speed (via $h$) and the local rung dynamics
(via $d$), a combination that is not generally
available in uniform one-dimensional systems.
The selective field thus provides a physically
transparent and experimentally accessible control
knob that is unique to the ladder geometry.

From an experimental perspective, the two-leg ladder
geometry maps naturally onto two leading qubit platforms.
In \emph{superconducting transmon arrays}, each rung
corresponds to a pair of frequency-tuneable transmons
coupled via a tunable coupler; the selective field $h$
is implemented by detuning the mediating transmon pair
away from the terminal pairs, which is a standard
single-qubit operation~\cite{krantz2019quantum}.
For $J/2\pi \sim 10$~MHz the fast oscillation period
is $T_{\rm fast} \approx 40$~ns and the slow transfer
time at $h/J = 100$ is $T_{\rm slow} \approx 240$~ns,
both well within current $T_2^* \gtrsim 50\,\mu$s
coherence times~\cite{liang2024enhanced}.
In \emph{semiconductor quantum-dot spin chains}, the
rung pairs correspond to singlet-triplet qubits and
the selective field maps to a local electrostatic gate
voltage that shifts the exchange coupling on the
mediating dots~\cite{chatterjee2021semiconductor,
baeumer2025perfect}. The robustness to $\delta \leq 10\%$
coupling disorder demonstrated in Sec.~\ref{sec:disorder}
is directly relevant to the fabrication tolerances of
current quantum-dot devices.

% ============================================================
\section{Effect of System Parameters on Transfer}
\label{sec:parameters}
% ============================================================

\subsection{Magnetic field strength}
\label{subsec:field}

We vary $h$ while keeping all other parameters fixed.

At $h = 10$ (Supplementary Fig.~S1), the separation
between $J$ and $h$ is reduced. The effective channel
picture begins to break down: wavelets appear on the
peaks of $C_{12}$ and $C_{56}$, indicating partial
re-absorption of entanglement by the mediating rung
($C_{34}$ develops non-negligible amplitude).
The maximum of $C_{56}$ decreases slightly below unity.

At $h = 0$, all three concurrences evolve in a complex,
non-periodic pattern and the transfer objective fails.

When $h$ is applied uniformly, peak $C_{56}$ drops to
$\sim0.8$, confirming that \emph{selectivity} ---
restriction to the mediating rungs --- is essential.

\subsection{Anisotropy parameters $d$ and $g$}
\label{subsec:anisotropy}

Setting $d = 0$ yields the $XY$ Hamiltonian. The initial
pair's entanglement decays gradually, and $C_{56}$ exhibits
a nearly periodic but decaying envelope. The Ising coupling
thus plays an important stabilising role in preserving the
transfer amplitude.

Setting $g = 0$ transforms the Hamiltonian into the
$XXZ$ model. The initial pair's entanglement decays
slowly below unity, while $C_{34}$ and $C_{56}$ remain
close to zero. This arises because the initial state
$|\Phi^+\rangle$ is close to an eigenstate of the $XXZ$
Hamiltonian in this parameter regime, suppressing
the dynamics required for transfer.

\subsection{Parameter Space Survey}
\label{subsec:heatmap}

\begin{figure}[tb]
    \centering
    \includegraphics[width=\columnwidth]{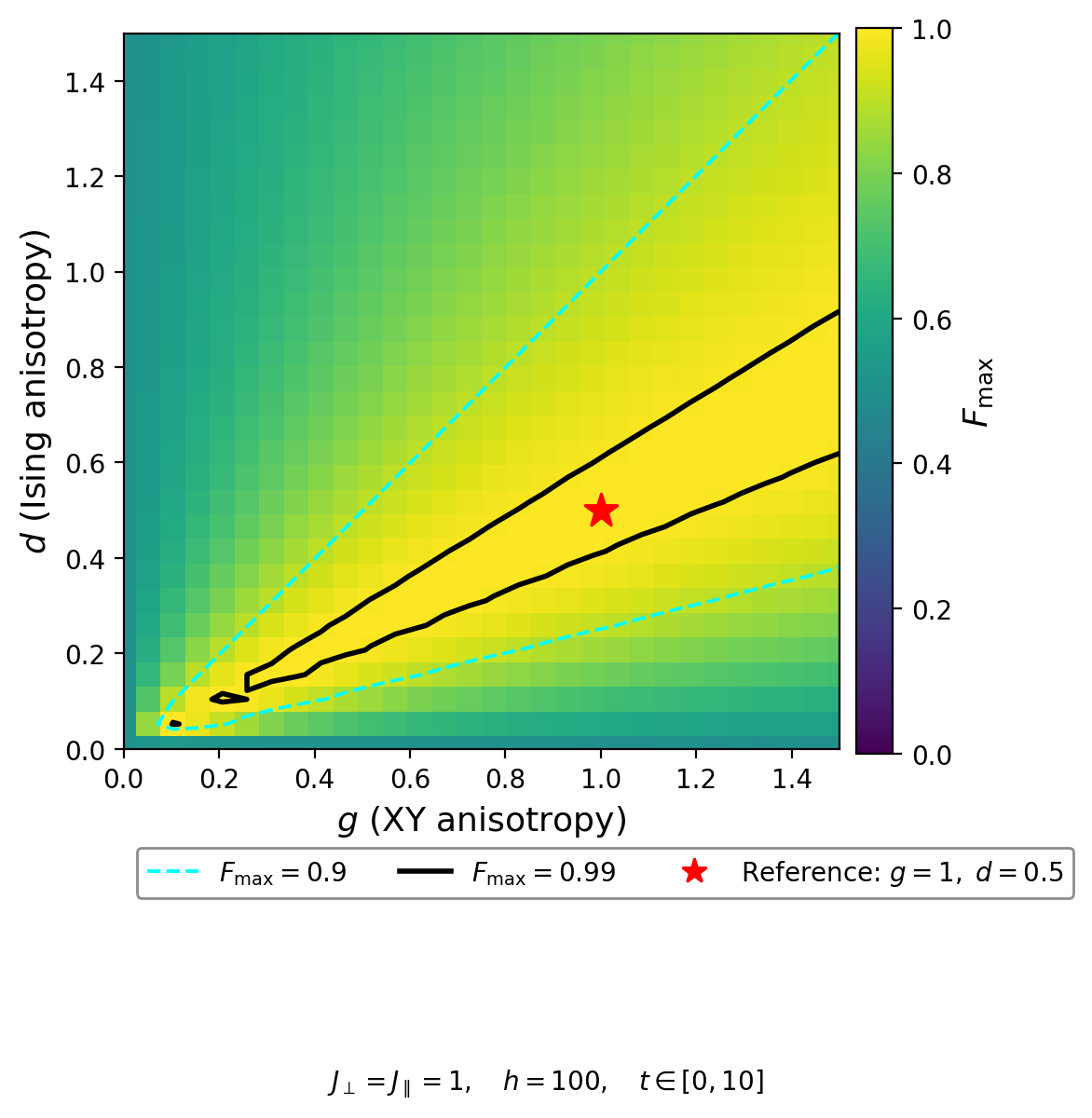}
    \caption{Peak fidelity $F_{\max}(g,d)$ for
    $J_\perp=J_\parallel=1$, $h=100$, $t\in[0,10]$.
    Black contours: $F_{\max}=0.99$. Cyan dashed:
    $F_{\max}=0.9$. Red star: reference parameters
    ($g=1$, $d=0.5$). Both axes at $g=0$ and $d=0$
    show low fidelity; the diagonal band shows that
    both anisotropy parameters must be non-zero.}
    \label{fig:heatmap}
\end{figure}

To map the global dependence on anisotropy parameters,
we compute $F_{\max}$ over a $30\times30$ grid of
$(g,d)$ values ($J=1$, $h=100$, $t\in[0,10]$).
Figure~\ref{fig:heatmap} shows a clear diagonal
high-fidelity band ($F_{\max}>0.99$, black contours):
both $g$ and $d$ must be non-zero simultaneously.
The reference parameters (red star) sit well inside
the optimal region with $F_{\max}=0.9998$. The ratio
$d/g$ is the key control parameter, with optimal
transfer for $0.3\lesssim d/g\lesssim0.7$.

% ============================================================
\section{Effect of Initial State}
\label{sec:initialstates}
% ============================================================

We now vary the initial Bell state of the first pair,
keeping all other parameters at their reference values.
For all Bell states studied, $C_{34}$ remains close to
zero throughout, confirming the effective channel picture
is robust to the initial state choice. $C_{56}$ reaches
values close to unity periodically for all inputs, though
the oscillation pattern of $C_{12}$ changes. We show
the most physically distinct case: a fully separable
initial state.

\subsubsection{Separable initial state ($|00\rangle$)}

For a fully separable initial state $|00\rangle^{1,2}$
with $h = 100$, the entanglement profiles of the
first and last pairs resemble the reference case but
\emph{without} the time delay, because no entanglement
needs to be ``transferred'' --- the field generates
it directly in both terminal pairs simultaneously.
As $h$ is reduced, the plots of $C_{12}$ and $C_{56}$
become identical (as expected by symmetry), but with
degraded amplitude.

% ============================================================
\section{Robustness Under Coupling Disorder}
\label{sec:disorder}
% ============================================================

In any realistic physical implementation, the coupling
constants of the Hamiltonian will deviate from their
ideal values due to fabrication imperfections or
environmental fluctuations. It is therefore essential
to assess whether the high-fidelity entanglement transfer
demonstrated in the previous sections persists in the
presence of such disorder.

We model disorder by replacing each coupling constant
$J$ with $J(1 + \delta_k)$, where $\delta_k$ is drawn
independently and uniformly from the interval
$[-\delta, +\delta]$, and $\delta$ is the disorder
strength. This applies independently to each rung
coupling $J_\perp$ and each leg coupling $J_\parallel$,
while the selective magnetic field $h$ is kept fixed
(as it corresponds to an externally controlled parameter
rather than a material property).

For each disorder strength $\delta \in
\{0\%, 5\%, 10\%, 20\%\}$, we perform $N_{\rm samples}
= 200$ independent disorder realisations and compute
the disorder-averaged transfer fidelity
$\langle F(t) \rangle$ and the mean peak fidelity
$\langle F_{\max} \rangle$, defined as the average
of the maximum fidelity achieved over $t \in [0, 10]$
across all realisations.

\begin{figure*}[tb]
    \centering
    \includegraphics[width=\textwidth]{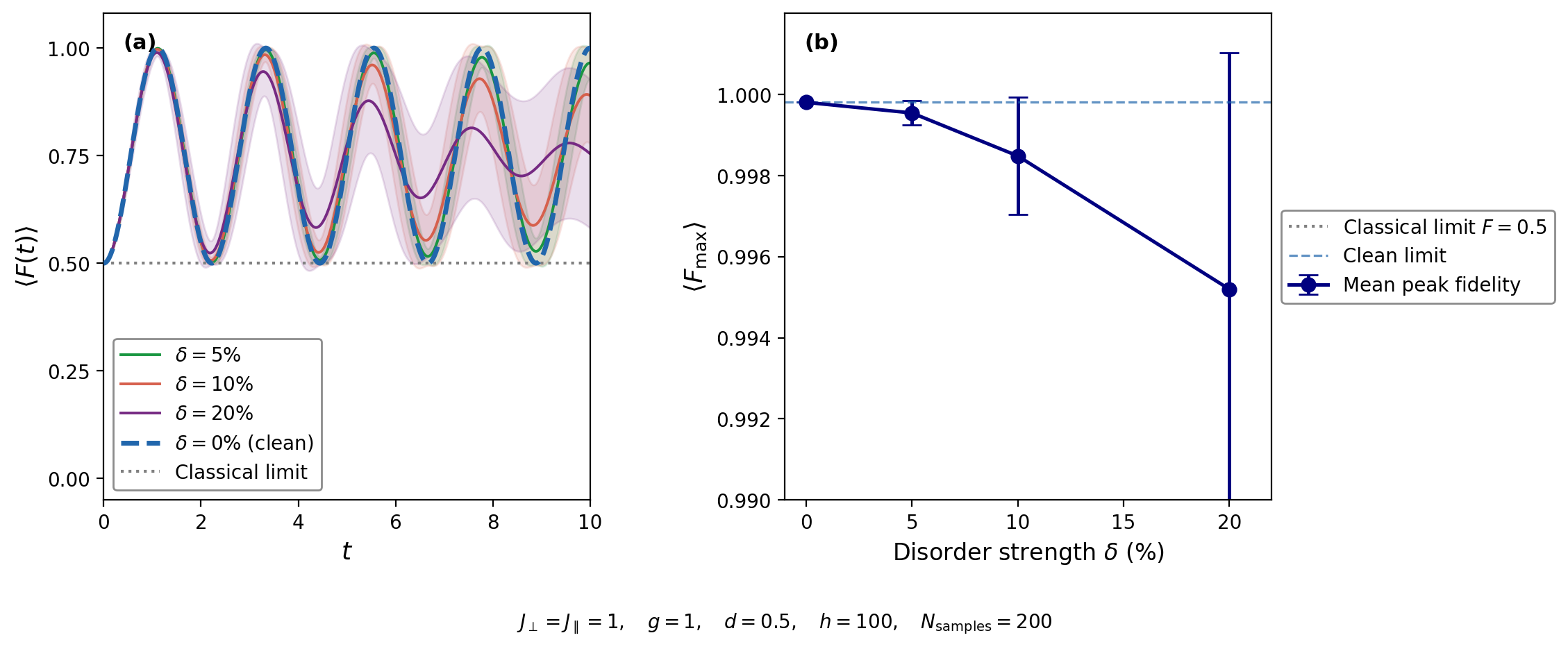}
    \caption{Robustness of entanglement transfer under
    coupling disorder, for reference parameters
    ($J_\perp = J_\parallel = 1$, $g = 1$, $d = 0.5$,
    $h = 100$) and $N_{\rm samples} = 200$ disorder
    realisations per disorder level.
    (a) Disorder-averaged transfer fidelity
    $\langle F(t) \rangle$ as a function of time
    for disorder strengths $\delta = 0\%$ (clean,
    thick dashed blue), $5\%$ (green), $10\%$ (orange),
    and $20\%$ (purple). Shaded bands indicate
    $\pm 1$ standard deviation across realisations.
    The dotted line marks the classical limit $F = 0.5$.
    (b) Mean peak fidelity $\langle F_{\max} \rangle$
    versus disorder strength $\delta$, with error bars
    showing $\pm 1$ standard deviation. The dashed
    line shows the clean-case limit. The transfer
    remains highly faithful ($\langle F_{\max} \rangle
    > 0.999$) for $\delta \leq 5\%$ and exceeds the
    classical limit for all disorder levels studied.}
    \label{fig:disorder}
\end{figure*}

Figure~\ref{fig:disorder} shows the results.
Panel~(a) displays the disorder-averaged fidelity
$\langle F(t) \rangle$ as a function of time for
each disorder level. The shaded bands indicate
$\pm 1$ standard deviation across the 200 realisations.
Panel~(b) shows the mean peak fidelity
$\langle F_{\max} \rangle$ as a function of disorder
strength $\delta$.

The key findings are:
\begin{enumerate}
  \item For weak disorder ($\delta = 5\%$), the
    disorder-averaged fidelity is virtually
    indistinguishable from the clean case, with
    $\langle F_{\max} \rangle = 0.9995 \pm 0.0003$.
    This level of disorder is typical of current
    state-of-the-art superconducting qubit
    fabrication~\cite{baeumer2025perfect}.
  \item For moderate disorder ($\delta = 10\%$),
    the transfer remains highly faithful, with
    $\langle F_{\max} \rangle = 0.9985 \pm 0.0014$
    --- a degradation of less than $0.2\%$ relative
    to the clean case.
  \item Even for strong disorder ($\delta = 20\%$),
    the mean peak fidelity remains
    $\langle F_{\max} \rangle = 0.9952 \pm 0.0058$,
    well above the classical limit of $F = 0.5$.
    The oscillation pattern in panel~(a) shows
    increased phase spread across realisations
    (wider shaded bands) but the peak transfer
    quality is remarkably well preserved.
  \item The fidelity never drops below the classical
    limit $F = 0.5$ for any disorder level studied,
    confirming that the transfer remains genuinely
    quantum for all realistic imperfection levels.
\end{enumerate}

The robustness of the transfer can be understood
physically: the dominant mechanism is the effective
coupling $J_{\rm eff} = \alpha(d,g)\,J^2/h$ between
the terminal pairs, mediated by virtual excitations
through the frozen mediating rung. Since $h \gg J$,
small variations in $J$ produce only second-order
changes in $J_{\rm eff}$, making the transfer
inherently robust to first-order coupling fluctuations.

We note that our disorder model --- independent random
fluctuations of each coupling constant --- represents
the most generic case of fabrication imperfections.
This differs from the correlated disorder studied
by Almeida \emph{et al.}~\cite{almeida2019disordered}
in a two-leg ladder context, where long-range
correlated disorder was found to enhance transfer
fidelity by suppressing Anderson localisation. Our
results provide the complementary baseline: even
under uncorrelated disorder, the transfer fidelity
remains high for $\delta \lesssim 10\%$. We note
that the present disorder model addresses coupling
fluctuations only; inhomogeneity in the applied
field $h$ may have a stronger effect, since $h$
directly sets the energy gap that governs the virtual
coupling mechanism. This sensitivity to field
fluctuations is left for future investigation.

% ============================================================
\section{Scaling to Larger Systems}
\label{sec:scaling}
% ============================================================

A natural question --- raised explicitly by referee
reports on earlier versions of this work --- is whether
the transfer mechanism persists as the number of rung
pairs $N$ increases beyond the minimal three-pair system.

\begin{widetext}
\begin{center}
\noindent\parbox{\textwidth}{%
\includegraphics[width=\textwidth,height=0.28\textheight,keepaspectratio]{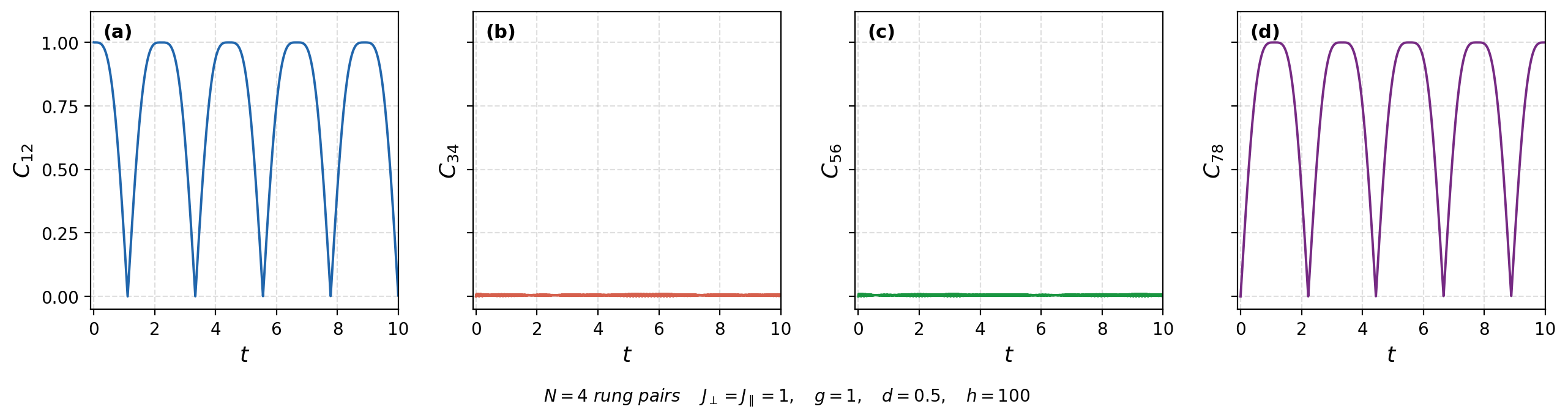}\\
\includegraphics[width=\textwidth,height=0.28\textheight,keepaspectratio]{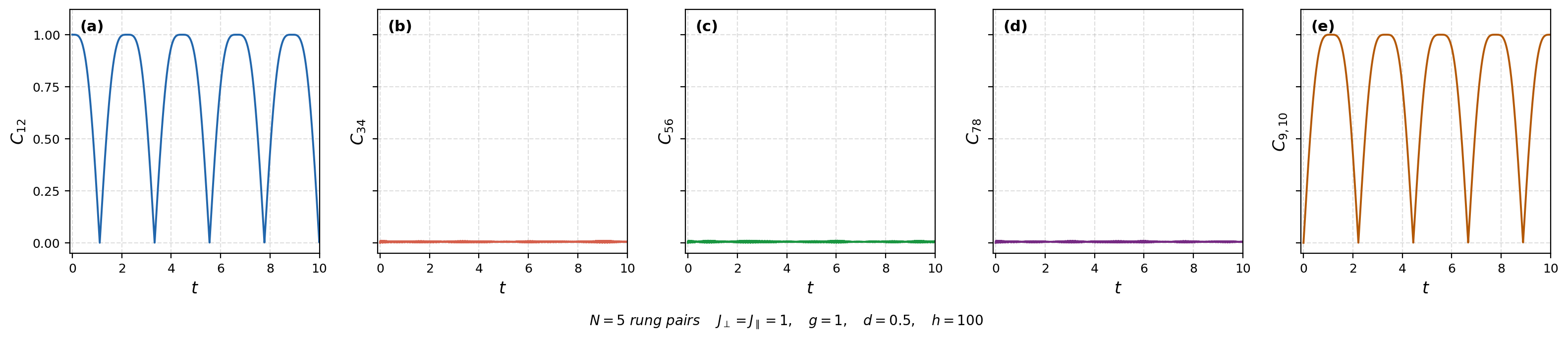}%
}
\captionof*{figure}{FIG.~7. Scaling of the entanglement transfer mechanism
    to larger system sizes, with reference parameters
    ($J_\perp = J_\parallel = 1$, $g = 1$, $d = 0.5$,
    $h = 100$).
    \textbf{Top row} ($N=4$, 8 sites, initial state
    $|\Phi^+\rangle^{1,2}\otimes|0\rangle^{\otimes 6}$):
    $C_{12}$, $C_{34}$, $C_{56}$, and $C_{78}$ vs time.
    The terminal pair $C_{78}$ oscillates in antiphase
    with $C_{12}$, while both mediating pairs remain
    close to zero.
    \textbf{Bottom row} ($N=5$, 10 sites, initial state
    $|\Phi^+\rangle^{1,2}\otimes|0\rangle^{\otimes 8}$):
    $C_{12}$, $C_{34}$, $C_{56}$, $C_{78}$, and
    $C_{9,10}$ vs time. All three mediating pairs
    remain close to zero, and $C_{9,10}$ oscillates
    in antiphase with $C_{12}$ with a longer period
    than the $N=4$ case, consistent with the longer
    effective path length.}
\label{fig:scaling}   
\end{center}
\end{widetext}
Figures~7 show results for
$N = 4$ and $N = 5$ rung pairs, using the reference
parameter set and the initial state of
Eq.~(\ref{eq:initial}).
The key observations are:
\begin{itemize}
  \item The antiphase oscillation between $C_{12}$ and
    $C_{N(N-1)}$ (the terminal pair) persists for all
    system sizes studied.
  \item The transfer period increases with $N$, consistent
    with the longer effective path length.
  \item The mediating pairs remain close to zero throughout,
    confirming the scalability of the effective channel
    picture.
  \item The maximum concurrence of the terminal pair
    decreases slightly with increasing $N$, indicating
    that larger systems require longer times and
    exhibit somewhat reduced peak fidelity.
\end{itemize}

The present study is limited to $N \leq 5$ rung pairs,
corresponding to a maximum Hilbert space dimension of
$2^{10} = 1024$. The computational cost of exact
diagonalisation scales as $2^{2N} \times 2^{2N}$,
making the $N = 6$ case ($4096 \times 4096$ matrices)
numerically demanding for the dense time-evolution
approach employed here. Extension to larger system
sizes would benefit from matrix product state or
density matrix renormalization group (DMRG) methods,
which have been successfully applied to multi-leg
Heisenberg ladders with much larger system
sizes~\cite{Li2024}. An alternative approach to
accessing larger system sizes is through block
renormalization group methods, which have been shown
to preserve essential physical features --- including
entanglement entropy and correlation functions --- of
one-dimensional spin Hamiltonians with a systematically
reduced number of spins~\cite{Mehrabankar2024}.
Whether such methods can be adapted to the two-leg
ladder geometry studied here remains an open question.

We emphasise, however, that the present results do not
establish asymptotic scaling behaviour. In particular,
the dependence of peak fidelity on system size may be
affected by dispersion and higher-order virtual
processes as $N$ increases. The consistent behaviour
observed across $N = 3$, $4$, and $5$ --- clean
antiphase oscillation between terminal pairs with all
mediating pairs remaining close to zero --- provides
numerical evidence, for small system sizes ($N \leq 5$),
that the transfer mechanism persists beyond the minimal
three-pair system. A definitive large-$N$ analysis is
left for future work; whether the transfer mechanism
persists in the thermodynamic limit remains an open
question.

% ============================================================
\section{Conclusion}
\label{sec:conclusion}
% ============================================================

We have studied entanglement transfer dynamics in a
two-leg spin-$\frac{1}{2}$ Heisenberg ladder with a
selective magnetic field, using exact diagonalisation
for $N\leq5$ rung pairs. The key contributions of
this work are:

The ladder supports high-fidelity Bell-state transfer
($F_{\max}=0.9998$ for $N=3$) when the selective field
is strong ($h\gg J$), the initial pair is in $|\Phi^+\rangle$,
and the anisotropy satisfies $g=1$, $d=0.5$. We have
analytically characterised the dynamics as a
\emph{two-scale quantum beat}: a fast carrier at
$h$-independent frequency $\omega_{\rm fast}=
2\sqrt{1+4d^2}\,J$ (verified to $0.1\%$,
Table~\ref{tab:gap}), and a slow transfer envelope
governed by the virtual inter-rung coupling
$J_{\rm eff} = \alpha(d,g)\,J^2/h$, derived from
second-order perturbation theory via two parallel
virtual rail paths (Eq.~\ref{eq:Heff}).

The $XY$ anisotropy $g$ and Ising anisotropy $d$
must be simultaneously non-zero: the global parameter
space map (Fig.~\ref{fig:heatmap}) shows a diagonal
high-fidelity band with $d/g$ as the key control
parameter. The transfer is robust to coupling disorder:
$\langle F_{\max}\rangle>0.998$ for $\delta\leq10\%$.
The mechanism scales to $N=4$ and $N=5$ with gradually
decreasing peak fidelity; extension to larger $N$
requires tensor-network methods.

The results are relevant to current experimental
platforms. For superconducting transmon arrays with
$J/2\pi\sim10$~MHz~\cite{krantz2019quantum,liang2024enhanced},
the fast oscillation period is $T_{\rm fast}\approx40$~ns
and the slow transfer time at $h/J=100$ is
$T_{\rm slow}\approx240$~ns, well within current
transmon coherence times ($T_2^* \gtrsim 50\,\mu$s)
and quantum-dot coherence times
($T_2^* \gtrsim 100\,\mu$s)~\cite{chatterjee2021semiconductor}.
Semiconductor quantum-dot
spin chains~\cite{chatterjee2021semiconductor,baeumer2025perfect}
provide an alternative platform.
Future directions include decoherence modelling via
Lindblad master equations, large-$N$ extension via
DMRG, and optimisation of the speed-fidelity trade-off
through non-uniform field profiles.

\vspace{0.5cm}
\begin{acknowledgments}
Authors S.G.\ and A.S.\ gratefully acknowledge
financial support from Islamic Azad University,
Omidiyeh Branch.
S.M.\ acknowledges support from the Australian Research
Council and the Queensland Quantum and Advanced
Technologies Institute, Griffith University.
\end{acknowledgments}

% ============================================================
%  BIBLIOGRAPHY
% ============================================================

\end{document}

% ============================================================

\title{Supplementary Information:\\
Entanglement Transfer Dynamics in a Two-Leg Spin Ladder
Under a Selective Magnetic Field}

\author{Soghra Ghanavati}
\affiliation{Department of Physics, Omidiyeh Branch,
  Islamic Azad University, Omidiyeh, Iran}

\author{Abbas Sabour}
\affiliation{Department of Physics, Omidiyeh Branch,
  Islamic Azad University, Omidiyeh, Iran}

\author{Somayeh Mehrabankar}
\email{somayeh.mehrabankar@yahoo.com}
\affiliation{Queensland Quantum and Advanced Technologies Institute,
  Griffith University, Yuggera Country, Brisbane, QLD 4111, Australia}

\date{March 29, 2026}

\maketitle

This Supplementary Information provides additional figures
supporting the results presented in the main text. The
figures are organised into three groups: (S1) the role of
magnetic field strength and selectivity, (S2) the effect
of anisotropy parameters, and (S3) dependence on the
initial state and quantum mutual information. All
parameters are as defined in the main text unless
otherwise stated.

% ============================================================
\section{Effect of Magnetic Field Strength and Selectivity}
\label{sec:SI_field}
% ============================================================

The main text discusses three cases: $h = 10$, $h = 0$,
and uniform field (applied to all sites). Here we provide
the corresponding time-series figures that confirm both
the \emph{strength} ($h \gg J$) and the \emph{selectivity}
(mediating rungs only) of the field are essential.

\begin{figure}[ht]
    \centering
    \includegraphics[width=\textwidth]{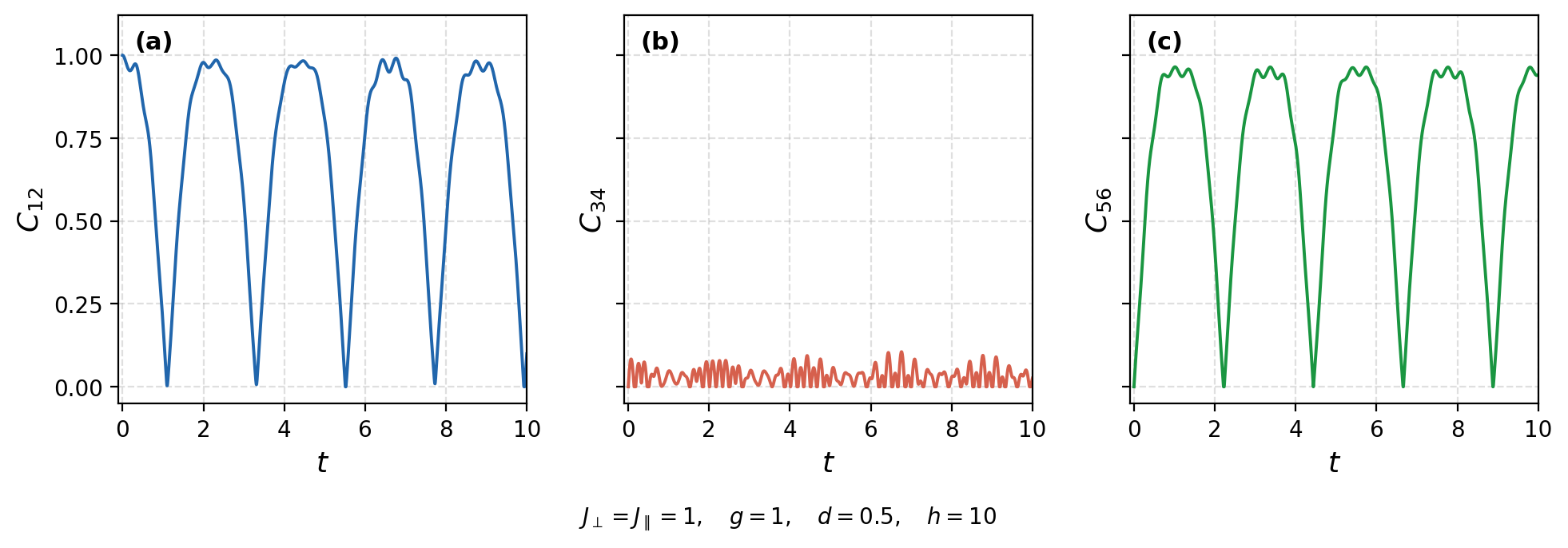}
    \caption{Time evolution of $C_{12}(t)$, $C_{34}(t)$,
    and $C_{56}(t)$ for reduced magnetic field $h = 10$,
    with reference parameters
    ($J_\perp = J_\parallel = 1$, $g = 1$, $d = 0.5$).
    (a) $C_{12}$ shows wavelet-like oscillations and no
    longer reaches unity cleanly.
    (b) $C_{34}$ develops non-negligible amplitude,
    indicating the effective channel picture begins to
    break down as $h$ approaches $J$.
    (c) $C_{56}$ exhibits wavelets at its peaks with
    maximum values slightly below unity.
    Reducing the field strength degrades transfer fidelity,
    confirming that $h \gg J$ is essential.}
    \label{fig:SI_h10}
\end{figure}

\begin{figure}[ht]
    \centering
    \includegraphics[width=\textwidth]{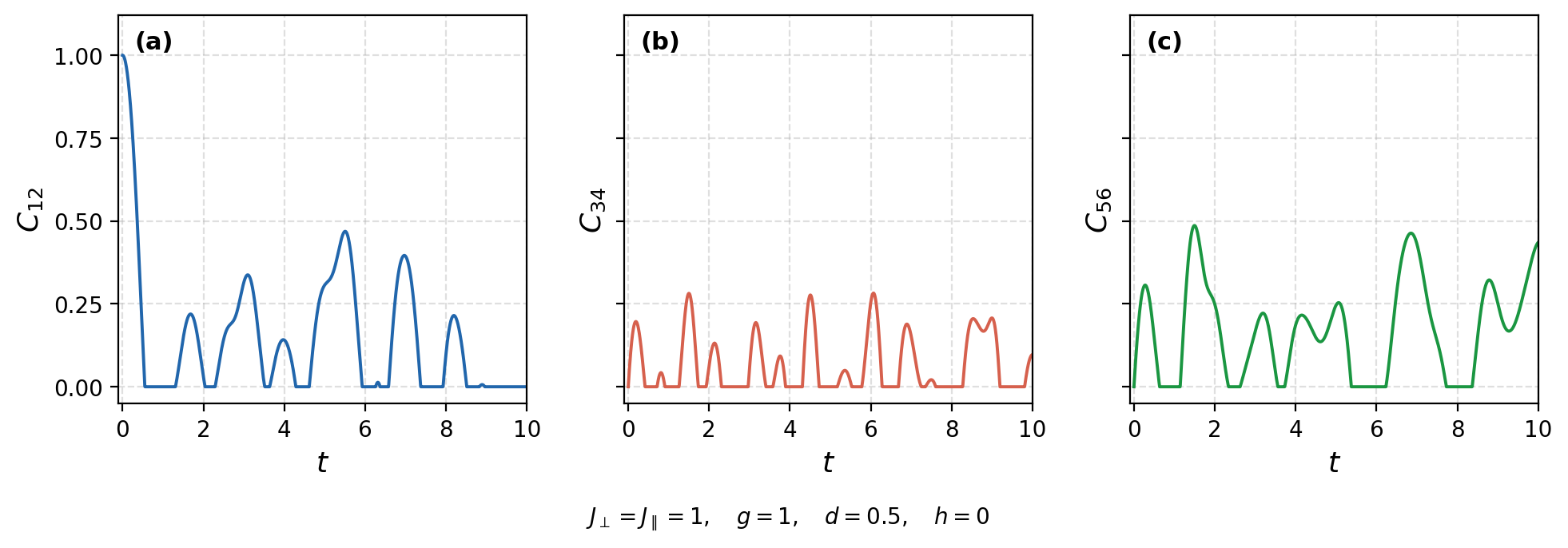}
    \caption{Time evolution of $C_{12}(t)$, $C_{34}(t)$,
    and $C_{56}(t)$ for zero magnetic field $h = 0$,
    with all other parameters at reference values
    ($J_\perp = J_\parallel = 1$, $g = 1$, $d = 0.5$).
    (a) $C_{12}$ decays rapidly and exhibits complex,
    non-periodic oscillations.
    (b) $C_{34}$ develops significant entanglement,
    showing the mediating rung actively participates
    in the dynamics.
    (c) $C_{56}$ reaches only modest values in an
    irregular, non-monotonic pattern.
    Without the selective field, the transfer objective
    fails completely, confirming that $h$ is an
    essential ingredient of the protocol.}
    \label{fig:SI_h0}
\end{figure}

\begin{figure}[ht]
    \centering
    \includegraphics[width=\textwidth]{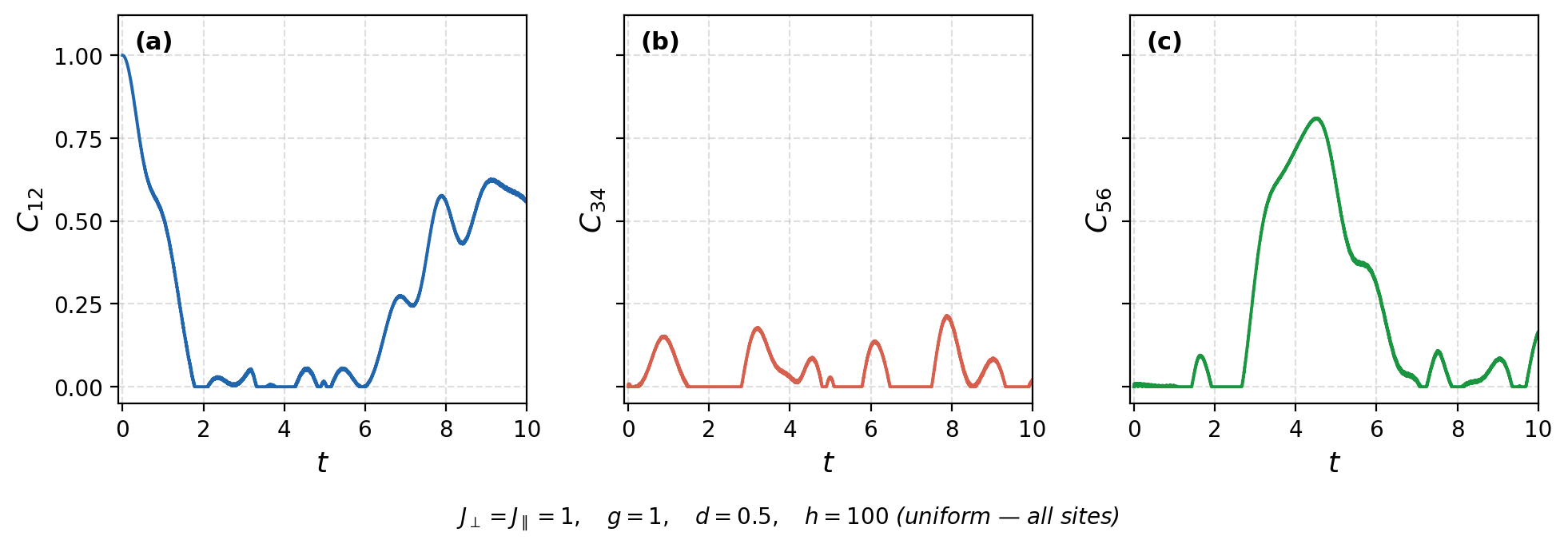}
    \caption{Time evolution of $C_{12}(t)$, $C_{34}(t)$,
    and $C_{56}(t)$ when the magnetic field $h = 100$
    is applied uniformly to \emph{all} sites, including
    the terminal rung pairs
    ($J_\perp = J_\parallel = 1$, $g = 1$, $d = 0.5$).
    (a) $C_{12}$ decays from unity and exhibits
    irregular behaviour.
    (b) $C_{34}$ develops intermittent bursts of
    entanglement.
    (c) $C_{56}$ reaches a maximum of only $\sim 0.8$,
    significantly below the high-fidelity transfer
    achieved with the selective field.
    The \emph{selectivity} of the magnetic field ---
    its restriction to the mediating rungs only ---
    is essential for high-fidelity transfer.}
    \label{fig:SI_huniform}
\end{figure}

% ============================================================
\section{Effect of Anisotropy Parameters}
\label{sec:SI_anisotropy}
% ============================================================

The main text presents only the $XY$ case ($d=0$,
Fig.~S1 of the main text) and discusses the $XXZ$
and pure $XX$ cases in text. Here we provide the
corresponding time-series figures for completeness.

\begin{figure}[ht]
    \centering
    \includegraphics[width=\textwidth]{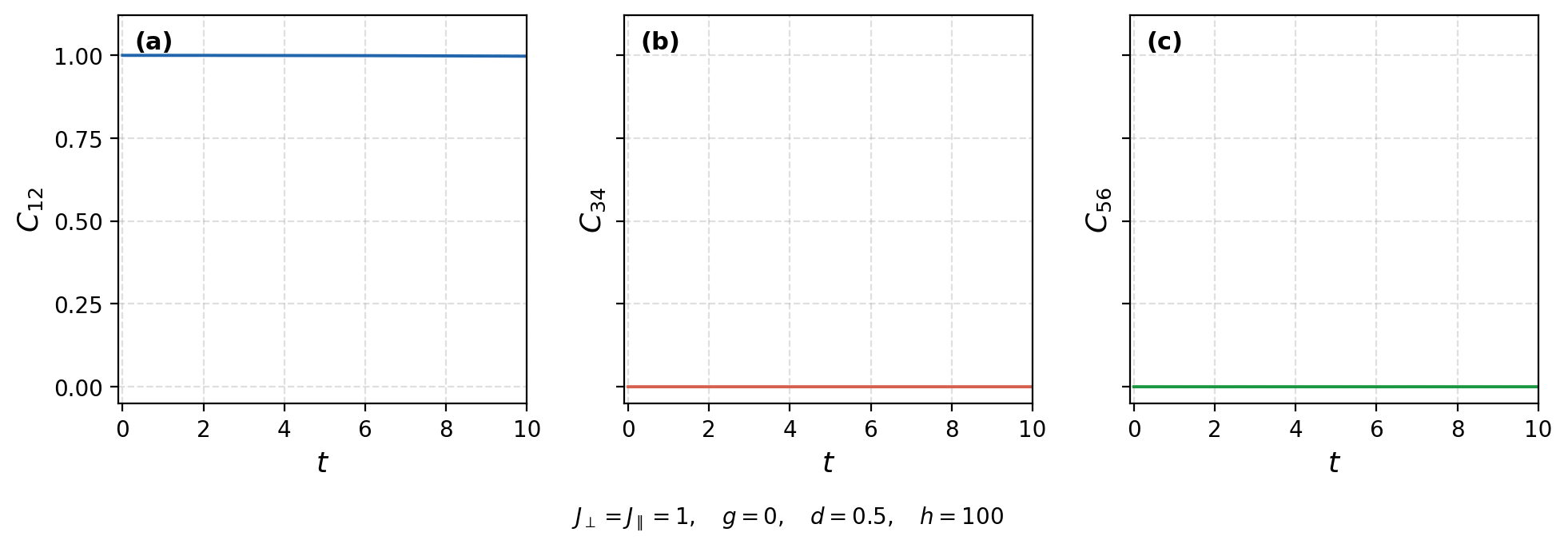}
    \caption{Time evolution of $C_{12}(t)$, $C_{34}(t)$,
    and $C_{56}(t)$ for the $XXZ$ Hamiltonian
    ($g = 0$, $d = 0.5$),
    with $J_\perp = J_\parallel = 1$ and $h = 100$.
    (a) $C_{12}$ remains constant at unity throughout,
    showing no decay or oscillation.
    (b) $C_{34}$ remains identically zero.
    (c) $C_{56}$ remains identically zero.
    The initial Bell state $|\Phi^+\rangle$ is close
    to an eigenstate of the $XXZ$ Hamiltonian at
    $g = 0$, suppressing all dynamics required for
    transfer. The $XY$ parameter $g$ is an essential
    ingredient of the protocol.}
    \label{fig:SI_g0}
\end{figure}

\begin{figure}[ht]
    \centering
    \includegraphics[width=\textwidth]{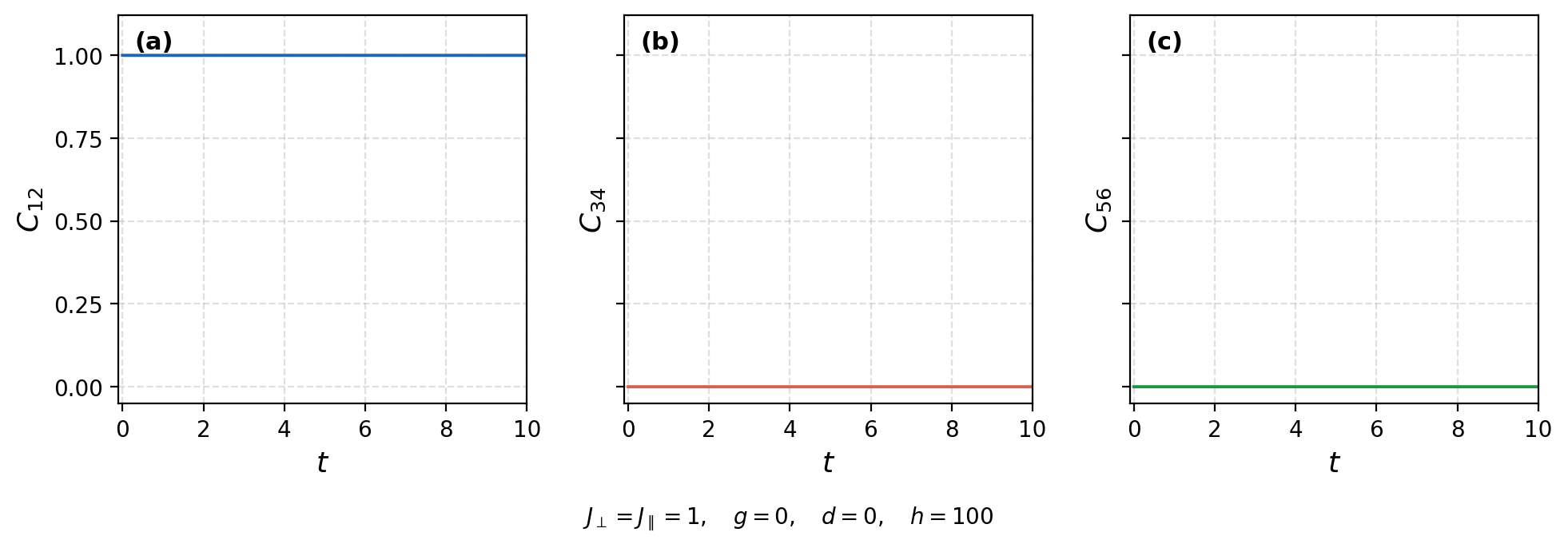}
    \caption{Time evolution of $C_{12}(t)$, $C_{34}(t)$,
    and $C_{56}(t)$ for the pure $XX$ Hamiltonian
    ($g = 0$, $d = 0$),
    with $J_\perp = J_\parallel = 1$ and $h = 100$.
    (a) $C_{12}$ remains constant at unity, identical
    to the $XXZ$ case of Fig.~\ref{fig:SI_g0}.
    (b) $C_{34}$ remains identically zero.
    (c) $C_{56}$ remains identically zero.
    When both anisotropy parameters vanish, the
    transfer objective fails completely. Comparing
    with Fig.~\ref{fig:SI_g0} and the $XY$ case in
    the main text, we conclude that both $g$ and $d$
    must be non-zero simultaneously to support
    coherent entanglement transfer through the ladder.}
    \label{fig:SI_g0d0}
\end{figure}

% ============================================================
\section{Dependence on Initial State and Quantum
         Mutual Information}
\label{sec:SI_initial}
% ============================================================

\subsection*{Initial state variation}

The main text discusses initial-state robustness in
text only. Here we provide the time-series figures
for two additional initial states: $|\Psi^+\rangle$
and a Bell-state superposition.

\begin{figure}[ht]
    \centering
    \includegraphics[width=\textwidth]{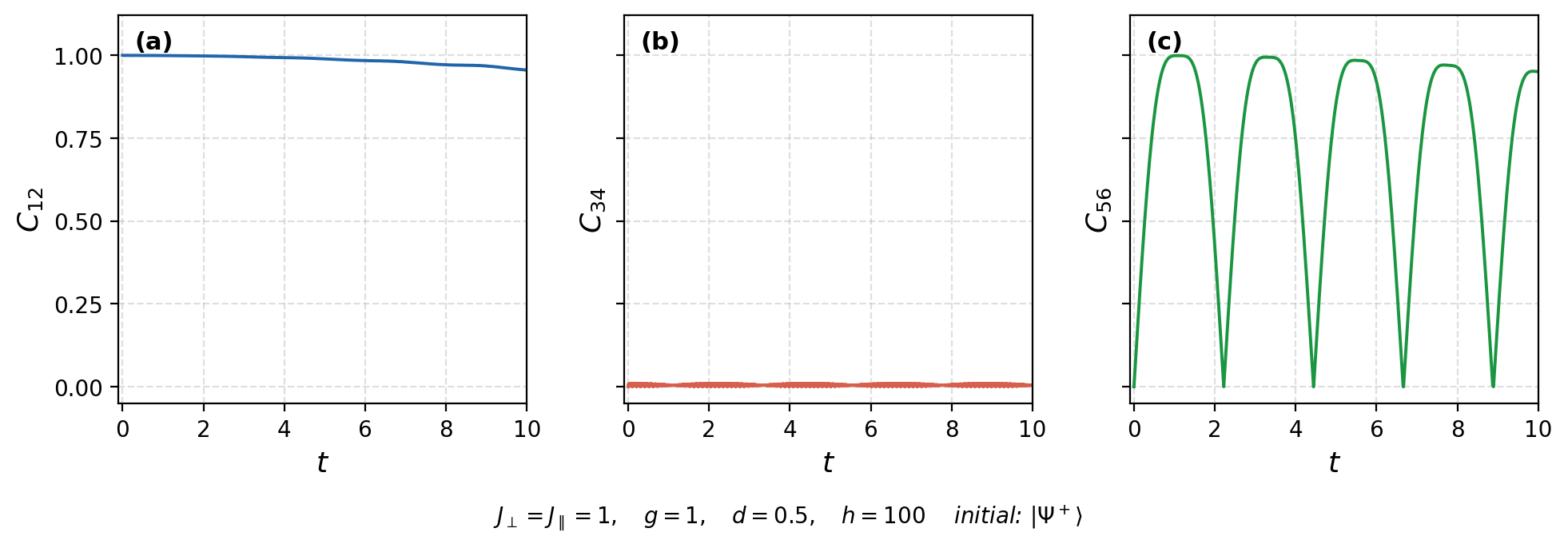}
    \caption{Time evolution of $C_{12}(t)$, $C_{34}(t)$,
    and $C_{56}(t)$ for initial Bell state
    $|\Psi^+\rangle^{1,2} = \frac{1}{\sqrt{2}}
    (|01\rangle + |10\rangle)$, with reference
    parameters ($J_\perp = J_\parallel = 1$, $g = 1$,
    $d = 0.5$, $h = 100$).
    (a) $C_{12}$ remains close to unity with only
    a very slow and gradual decay, and does not
    exhibit the periodic oscillation seen for
    $|\Phi^+\rangle$.
    (b) $C_{34}$ remains close to zero throughout,
    confirming the channel is transparent for all
    Bell state inputs.
    (c) $C_{56}$ oscillates and reaches values
    close to unity periodically. The alternate
    period of $C_{12}$ is significantly longer
    than that of $C_{56}$, reflecting the different
    symmetry of $|\Psi^+\rangle$ under the
    Hamiltonian.}
    \label{fig:SI_psiplus}
\end{figure}

\begin{figure}[ht]
    \centering
    \includegraphics[width=\textwidth]{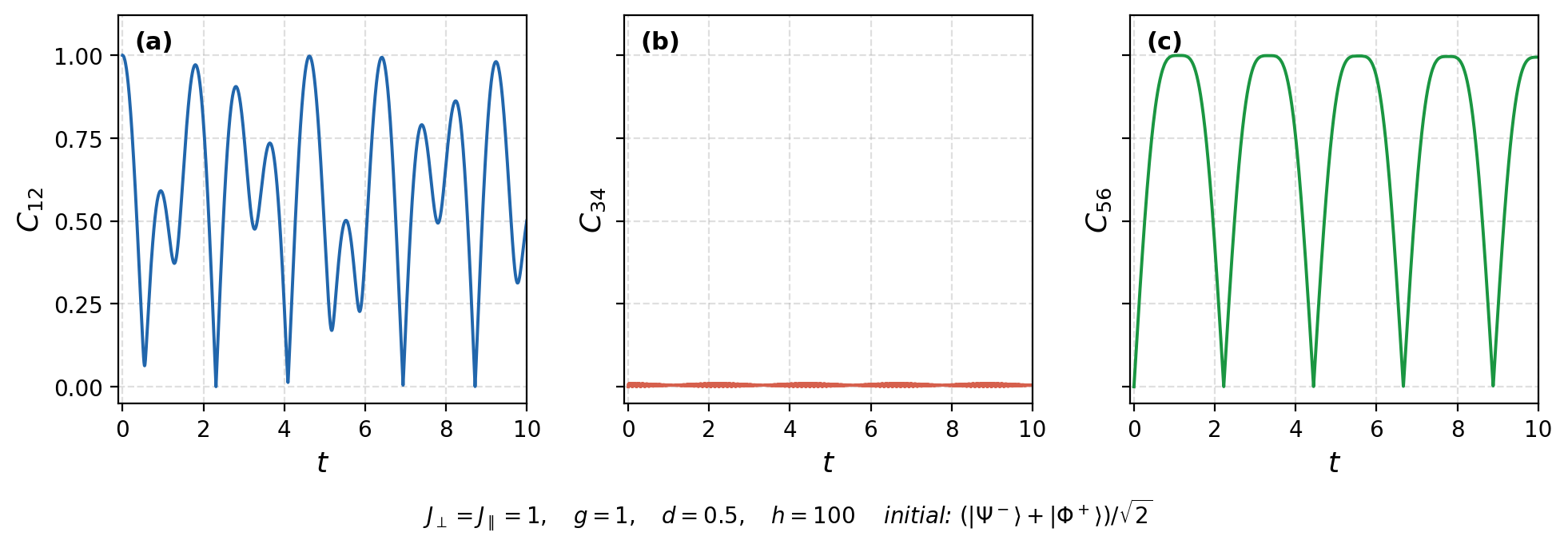}
    \caption{Time evolution of $C_{12}(t)$, $C_{34}(t)$,
    and $C_{56}(t)$ for the superposition initial
    state $\frac{1}{\sqrt{2}}(|\Psi^-\rangle +
    |\Phi^+\rangle)^{1,2}$, with reference parameters
    ($J_\perp = J_\parallel = 1$, $g = 1$,
    $d = 0.5$, $h = 100$).
    (a) $C_{12}$ exhibits complex, irregular
    oscillations reflecting the loss of definite
    symmetry under the Hamiltonian.
    (b) $C_{34}$ remains close to zero throughout,
    confirming the transparent channel behaviour
    is robust to the initial state.
    (c) $C_{56}$ continues to reach values close
    to unity periodically, demonstrating that
    entanglement transfer persists even when the
    initial pair dynamics are irregular.}
    \label{fig:SI_superpos}
\end{figure}

\subsection*{Quantum mutual information}

\begin{figure}[ht]
    \centering
    \includegraphics[width=\textwidth]{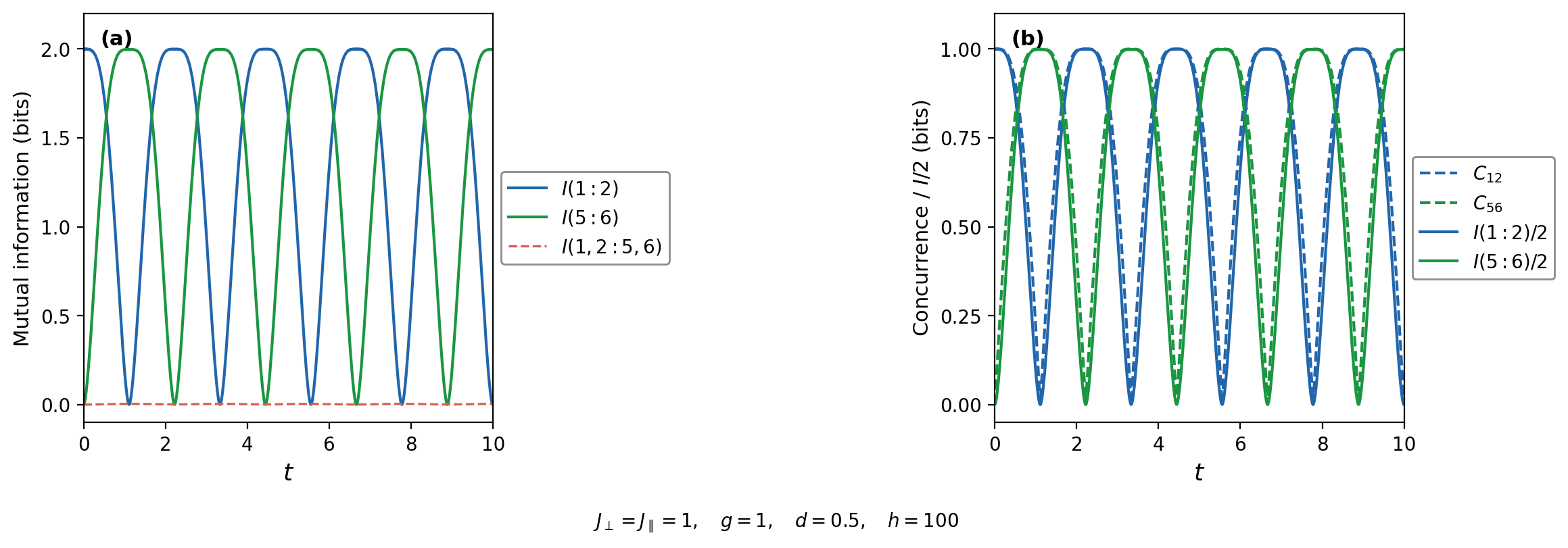}
    \caption{Quantum mutual information during
    entanglement transfer, for reference parameters
    ($J_\perp = J_\parallel = 1$, $g = 1$,
    $d = 0.5$, $h = 100$).
    (a) $I(1:2)$ (blue) and $I(5:6)$ (green) oscillate
    in perfect antiphase between 0 and 2 bits,
    confirming near-complete transfer of quantum
    information. $I(1,2:5,6)$ (red dashed) remains
    essentially zero throughout --- the two terminal
    pairs are never correlated with each other ---
    confirming the mediating rung acts as a truly
    transparent channel that routes entanglement
    without building up inter-pair correlations.
    (b) $I(i:j)/2$ (solid) vs concurrence $C_{ij}$
    (dashed) for both pairs. The near-perfect
    overlap confirms the two-qubit states remain
    close to pure states throughout, validating
    concurrence as a faithful proxy for mutual
    information in this system.
    The relationship $I \approx 2C$ holds to
    within $1\%$ for all times studied.}
    \label{fig:SI_MI}
\end{figure}

The quantum mutual information $I(A:B) = S(\rho_A) +
S(\rho_B) - S(\rho_{AB})$, where $S$ is the von Neumann
entropy, provides a complete characterisation of all
quantum and classical correlations. Figure~\ref{fig:SI_MI}
shows the time evolution of $I(1:2)$ and $I(5:6)$
(entanglement \emph{within} each terminal pair) and
$I(1,2:5,6)$ (correlations \emph{between} the two
terminal pairs).

